\documentclass[runningheads]{llncs}

\pdfoutput=1
\usepackage[hyphens]{url}
\usepackage[hyperindex,breaklinks]{hyperref}

\usepackage{graphicx}
\usepackage{booktabs}
\usepackage{textcomp}
\usepackage{amsmath}
\usepackage{caption}
\usepackage{subcaption}
\begin{document}
\title{Exploring End-to-End Techniques for Low-Resource Speech Recognition}
%
%
\author{Vladimir Bataev\inst{1} \and Maxim Korenevsky\inst{2,3} \and
Ivan Medennikov\inst{2,3} \and Alexander\\ Zatvornitskiy\inst{1,2,3}}
\authorrunning{V. Bataev et al.}
%
\institute{ Speech Technology Center Ltd, St.~Petersburg, Russia \and
STC-innovations Ltd, St.~Petersburg, Russia \and ITMO University, St.~Petersburg, Russia \\
\email{\{bataev,korenevsky,medennikov,zatvornitskiy\}@speechpro.com}}
\maketitle              
\begin{abstract}
In this work we present simple grapheme-based system for low-resource speech recognition using Babel data for Turkish spontaneous speech (80 hours). 
We have investigated different neural network architectures performance, including fully-convolutional, recurrent and ResNet with GRU. Different features and normalization techniques are compared as well. 
We also proposed CTC-loss modification using segmentation during training, which leads to improvement while decoding with small beam size.

Our best model achieved word error rate of 45.8\%, which is the best reported result for end-to-end systems using in-domain data for this task, according to our knowledge.

\end{abstract}

\section{Introduction}
Although development of the first speech recognition systems began half a century ago, there has been a significant increase of the accuracy of ASR systems and number of their applications for the recent ten years, even for low-resource languages~\cite{Levin14-ACC,Khomitsevich15-ABK}.

This is mainly due to widespread applying of deep learning and very effective performance of neural networks in hybrid recognition systems (DNN-HMM).
However, for last few years there has been a trend to change traditional ASR training paradigm. End-to-end training systems gradually displace complex multistage learning process (including training of GMMs \cite{StateLevelControl}, clustering of allophones’ states, aligning of speech to clustered senones, training neural networks with cross-entropy loss, followed by retraining with sequence-discriminative criterion). The new approach implies training the system in one global step, working only with acoustic data and reference texts, and significantly simplifies or even completely excludes in some cases the decoding process.
It also avoids the problem of out-of-vocabulary words (OOV), because end-to-end system, trained with parts of the words as targets, can construct new words itself using graphemes or subword units, while traditional DNN-HMM systems are limited with language model vocabulary.

The whole variety of end-to-end systems can be divided into 3 main categories: Connectionist Temporal Classification (CTC) \cite{Graves2006}; Sequence-to-sequence models with attention mechanism \cite{LAS2016}; RNN-Transducers \cite{Graves2012}.


{\bf Connectionist Temporal Classification} (CTC) approach uses loss functions that utilize all possible alignments between reference text and audio data. Targets for CTC-based system can be phonemes, graphemes, syllables and other subword units and even whole words. However, a lot more data is usually required to train such systems well, compared to traditional hybrid systems.

{\bf Sequence-to-sequence models} are used to map entire input sequences to output sequences without any assumptions about their alignment. The most popular architecture for sequence-to-sequence models is encoder-decoder model with attention. Encoder and decoder are usually constructed using recurrent neural networks, basic attention mechanism calculates energy weights that emphasize importance of encoder vectors for decoding on this step, and then sums all these vectors with energy weights. Encoder-decoder models with attention mechanism show results close to traditional DNN-HMM systems and in some cases surpass them, but for a number of reasons their usage is still rather limited. First of all, this is related to the fact, that such systems show best results when the duration of real utterances is close to the duration of utterances from training data. However, when the duration difference increases, the performance degrades significantly \cite[ Fig. 4 ``Utterance Length vs. Error'']{LAS2016}.

Moreover, the entire utterance must be preprocessed by encoder before start of decoder's work. This is the reason, why it is hard to apply the approach to recognize long recordings or streaming audio. Segmenting long recordings into shorter utterances solves the duration issue, but leads to a context break, and eventually negatively affects recognition accuracy.
Secondly, the computational complexity of encoder-decoder models is high because of recurrent networks usage, so these models are rather slow and hard to parallelize.

The idea of {\bf RNN-Transducer} is an extension of CTC and provides the ability to model inner dependencies separately and jointly between elements of both input (audio frames) and output (phonemes and other subword units) sequences. Despite of mathematical elegance, such systems are very complicated and hard to implement, so they are still rarely used, although several impressive results were obtained using this technique.

CTC-based approach is easier to implement, better scaled and has many ``degrees of freedom'', which allows to significantly improve baseline systems and achieve results close to state-of-the-art. Moreover, CTC-based systems are well compatible with traditional WFST-decoders and can be easily integrated with conventional ASR systems.

Besides, as already mentioned, CTC-systems are rather sensitive to the amount of training data, so it is very relevant to study how to build effective CTC-based recognition system using a small amount of training samples. It is especially actual for low-resource languages, where we have only a few dozen hours of speech. Building ASR system for low-resource languages is one of the aims of international Babel program, funded by the Intelligence Advanced Research Projects Activity (IARPA). Within the program extensive research was carried out, resulting in creation of a number of modern ASR systems for low-resource languages. Recently, end-to-end approaches were applied to this task, showing expectedly worse results than traditional systems, although the difference is rather small.

In this paper we explore a number of ways to improve end-to-end CTC-based systems in low-resource scenarios using the Turkish language dataset from the IARPA Babel collection. In the next section we describe in more details different versions of CTC-systems and their application for low-resource speech recognition.  Section 3 describes the experiments and their results. Section 4 summarizes the results and discusses possible ways for further work.

\section{Related work}

Development of CTC-based systems originates from the paper \cite{Graves2006} where CTC loss was introduced. This loss is a total probability of labels sequence given observation sequence, which takes into account all possible alignments induced by a given words sequence.

Although a number of possible alignments increases exponentially with sequences’ lengths, there is an efficient algorithm to compute CTC loss based on dynamic programming principle (known as Forward-Backward algorithm). This algorithm operates with posterior probabilities of any output sequence element observation given the time frame and CTC loss is differentiable with respect to these probabilities. 

Therefore, if an acoustic model is based on the neural network which estimates these posteriors, its training may be performed with a conventional error back-propagation gradient descent \cite{Rumelhart1988}. Training of ASR system based on such a model does not require an explicit alignment of input utterance to the elements of output sequence and thus may be performed in end-to-end fashion. It is also important that CTC loss accumulates the information about the whole output sequence, and hence its optimization is in some sense an alternative to the traditional fine-tuning of neural network acoustic models by means of sequence-discriminative criteria such as sMBR \cite{Kingsbury2009} etc. The implementation of CTC is conventionally based on RNN/LSTM networks, including bidirectional ones as acoustic models, since they are known to model long context effectively.

The important component of CTC is a special ``blank'' symbol which fills in gaps between meaningful elements of output sequence to equalize its length to the number of frames in the input sequence. It corresponds to a separate output neuron, and blank symbols are deleted from the recognized sequence to obtain the final result. In \cite{Wav2letter} a modification of CTC loss was proposed, referred as Auto SeGmentation criterion (ASG loss), which does not use blank symbols. Instead of using ``blank'', a simple transition probability model for an output symbols is introduced. This leads to a significant simplification and speedup of computations. Moreover, the improved recognition results compared to basic CTC loss were obtained.

DeepSpeech \cite{DeepSpeech2014} developed by Baidu Inc. was one of the first systems that demonstrated an effectiveness of CTC-based speech recognition in LVCSR tasks. Being trained on 2300 hours of English Conversational Telephone Speech data, it demonstrated state-of-the-art results on Hub5'00 evaluation set. Research in this direction continued and resulted in DeepSpeech2 architecture \cite{DeepSpeech2}, composed of both convolutional and recurrent layers. This system demonstrates improved accuracy of recognition of both English and Mandarin speech. Another successful example of applying CTC to LVCSR tasks is EESEN system \cite{EESEN}. It integrates an RNN-based model trained with CTC criterion to the conventional WFST-based decoder from the Kaldi toolkit \cite{Povey_ASRU2011}. The paper \cite{wav2Letter2} shows that end-to-end systems may be successfully built from convolutional layers only instead of recurrent ones. It was demonstrated that using Gated Convolutional Units  (GLU-CNNs) and training with ASG-loss leads to the state-of-the-art results on the LibriSpeech database (960 hours of training data).


Recently, a new modification of DeepSpeech2 architecture was proposed in \cite{Salesforce}. Several lower convolutional layers were replaced with a deep residual network with depth-wise separable convolutions. This modification along with using strong regularization and data augmentation techniques leads to the results close to DeepSpeech2 in spite of significantly lower amount of data used for training. Indeed, one of the models was trained with only 80 hours of speech data (which were augmented with noisy and speed-perturbed versions of original data).

\begin{figure*}[h]
\centering
\begin{subfigure}{.4\textwidth}
  \centering
  \includegraphics[width=.8\linewidth]{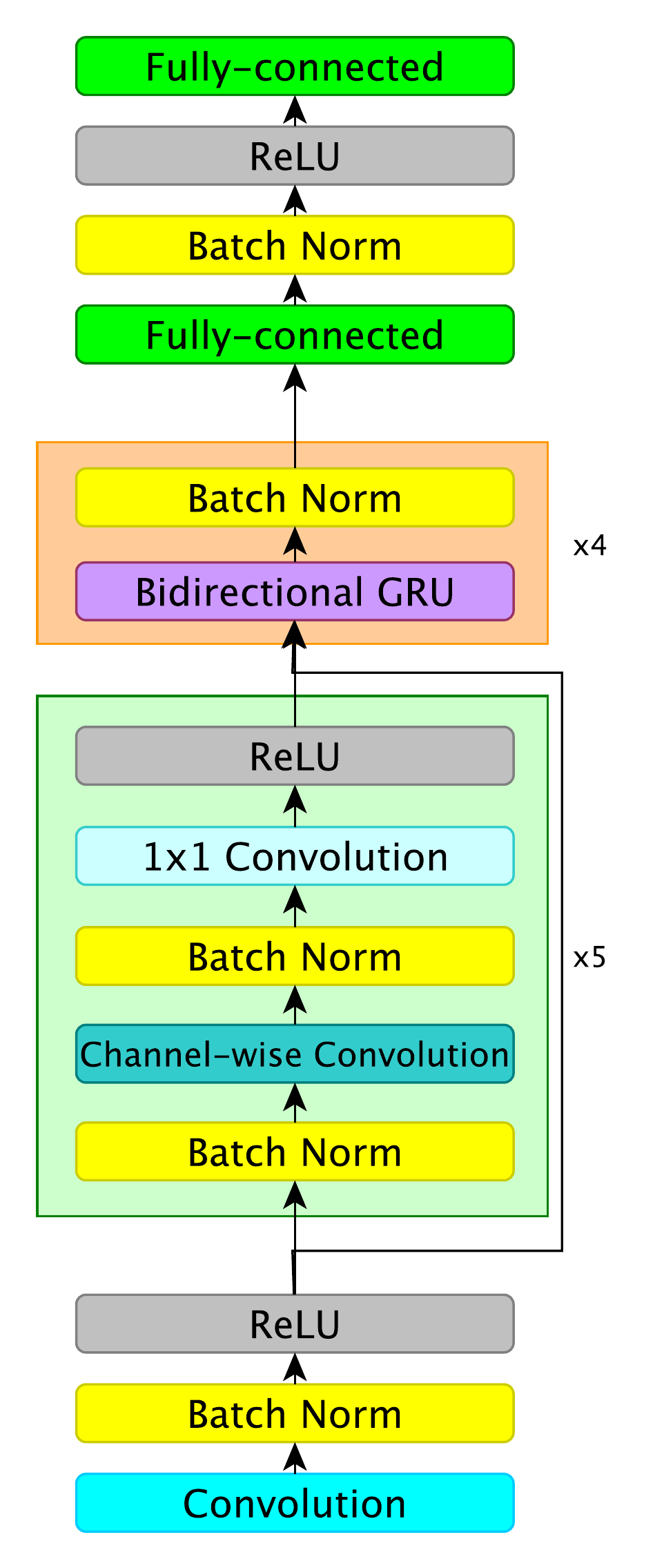}
  \caption{DeepSpeech-like model}
  \label{fig:sub1}
\end{subfigure}
\begin{subfigure}{.4\textwidth}
  \centering
  \includegraphics[width=.85\linewidth]{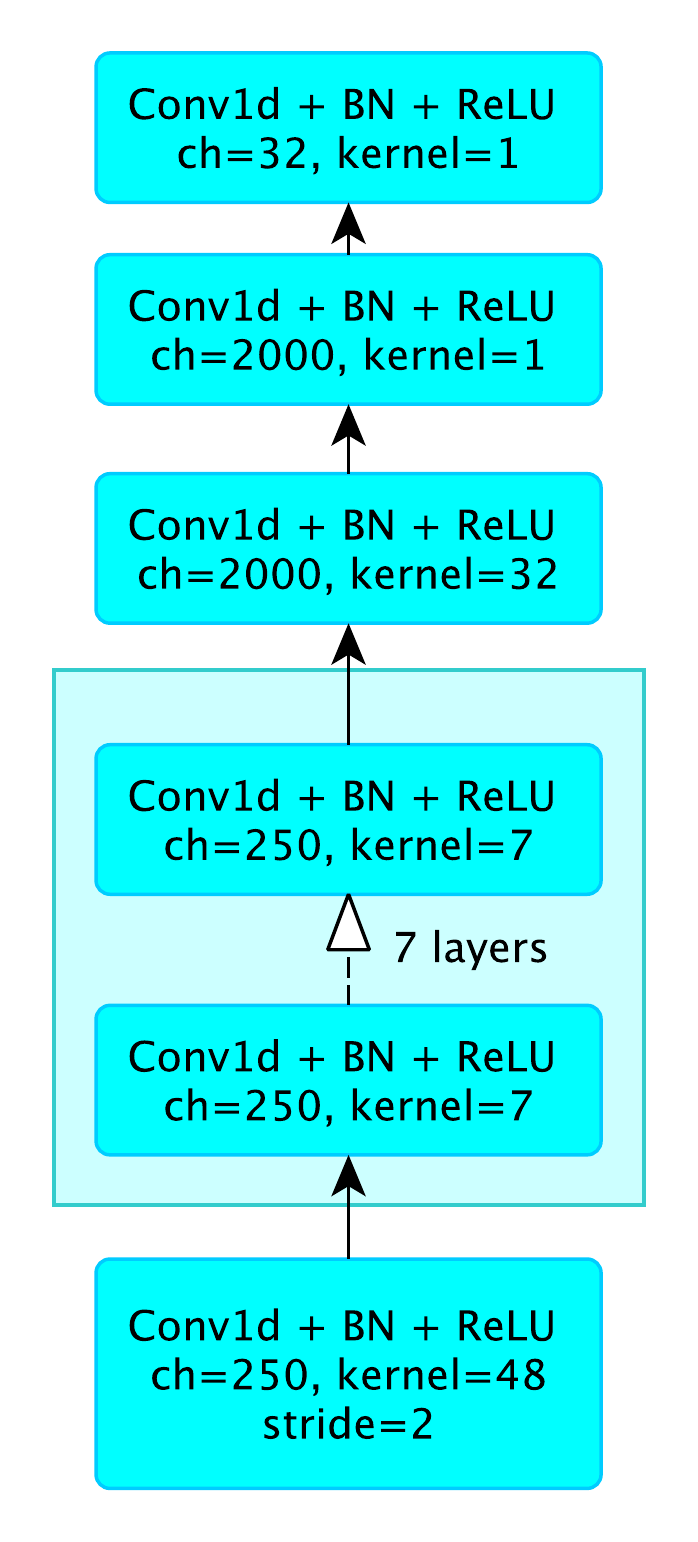}
  \caption{Wav2Letter}
  \label{fig:sub2}
\end{subfigure}
\caption{Architectures}
\label{fig:Architectures}
\end{figure*}


These results suggest that CTC can be successfully applied for the training of ASR systems for low-resource languages, in particular, for those included in Babel research program (the amount of training data for them is normally 40 to 80 hours of speech). 

Currently, Babel corpus contains data for more than 20 languages, and for most of them quite good traditional ASR system were built \cite{BBN,IBM,STC}. In order to improve speech recognition accuracy for a given language, data from other languages is widely used as well. It can be used to train multilingual system via multitask learning or to obtain high-level multilingual representations, usually bottleneck features, extracted from a pre-trained multilingual network.

One of the first attempts to build ASR system for low-resource BABEL languages using CTC-based end-to-end training was made recently \cite{Dalmia2018}. Despite the obtained results are somewhat worse compared to the state-of-the-art traditional systems, they still demonstrate that CTC-based approach is viable for building low-resource ASR systems. The aim of our work is to investigate some ways to improve the obtained results.

\section {Experiments}
\subsection{Basic setup}
For all experiments we used conversational speech from IARPA Babel Turkish Language Pack (LDC2016S10). This corpus contains about 80 hours of transcribed speech for training and 10 hours for development. The dataset is rather small compared to widely used benchmarks for conversational speech: English Switchboard corpus (300 hours, LDC97S62) and Fisher dataset (2000 hours, LDC2004S13 and LDC2005S13).

As targets we use 32 symbols: 29 lowercase characters of Turkish alphabet \cite{Alphabet}, apostrophe, space and special \textlangle blank\textrangle \space character that means ``no output''. Thus we do not use any prior linguistic knowledge and also avoid OOV problem as the system can construct new words directly. 

All models are trained with CTC-loss. Input features are 40 mel-scaled log filterbank enegries (FBanks) computed every 10 ms with 25 ms window, concatenated with deltas and delta-deltas (120 features in vector). We also tried to use spectrogram and experimented with different normalization techniques.

For decoding we used character-based beam search \cite{CTCDecoder} with 3-gram language model build with SRILM package \cite{SRILM} finding sequence of characters $c$ that maximizes the following objective \cite{DeepSpeech2014}:

\begin{equation*}
Q(c) = 	\log{P(c|x)} +\alpha\log{P_{lm}(c)} +\beta \text{wordcount}(c),
\end{equation*}
where $\alpha$ is language model weight and $\beta$ is word insertion penalty. 

For all experiments we used $\alpha = 0.8$, $\beta = 1$, and performed decoding with beam width equal to 100 and 2000, which is not very large compared to 7000 and more active hypotheses used in traditional WFST decoders (e.g. many Kaldi recipes do decoding with $max\_active = 7000$).

To compare with other published results \cite{Dalmia2018,KaldiTurResults} we used Sclite \cite{Sclite} scoring package to measure results of decoding with beam width 2000, that takes into account incomplete words and spoken noise in reference texts and doesn't penalize model if it incorrectly recognize these pieces. 

Also we report WER (word error rate) for simple argmax decoder (taking labels with maximum output on each time step and than applying CTC decoding rule – collapse repeated labels and remove ``blanks'').

\subsection{Experiments with architecture}

We tried to explore the behavior of different neural network architectures in case when rather small data is available. We used multi-layer bidirectional LSTM networks, tried fully-convolutional architecture similar to Wav2Letter \cite{Wav2letter} and explored DeepSpeech-like architecture developed by Salesforce (DS-SF) \cite{Salesforce}.

The convolutional model consists of 11 convolutional layers with batch normalization after each layer.
The DeepSpeech-like architecture consists of 5-layers residual network with depth-wise separable convolutions followed by 4-layer bidirectional Gated Recurrent Unit (GRU) as described in \cite{Salesforce}.

Our baseline bidirectional LSTM is 6-layers network with 320 hidden units per direction as in \cite{Dalmia2018}. Also we tried to use bLSTM to label every second frame (20 ms) concatenating every first output from first layer with second and taking this as input for second model layer.

The performance of our baseline models is shown in Table \ref{tab:Baseline_models}.

\begin{table}[h]
  \caption{Baseline models trained with CTC-loss}
  \label{tab:Baseline_models}
  \centering
    \begin{tabular}{ l c c c c c c }
    \toprule
    \textbf{Model} & \textbf{Step} & \textbf{Dropout} & \textbf{Argmax} & \multicolumn{2}{c}{\textbf{LM decoding}} & \textbf{Sclite} \\
    & & & & \textbf{beam 100} & \textbf{beam 2000} & \\ 
    \midrule
    Wav2Letter & 20ms & --- & 88.4 & 78.3 & 71.5 & 67.5 \\
    6-layer bLSTM & 10ms & --- & 69.9 & 61.1 & 56.3 & 51.7 \\
    6-layer bLSTM & 20ms & --- & 69.0 & 59.6 & \textbf{55.7} & \textbf{51.1} \\
    DS-SF & 20ms & no & 72.7 & 64.1 & 57.7 & 53.3 \\
    DS-SF & 20ms & between each layer & 71.8 & 59.41 & 55.7 & 50.8 \\
    DS-SF & 20ms & between modules & 68.6 & 58.9 & \textbf{54.5} & \textbf{49.7} \\
    \bottomrule
    \end{tabular}
\end{table}

\subsection{Loss modification: segmenting during training}

It is known that CTC-loss is very unstable for long utterances \cite{Graves2006}, and smaller utterances are more useful for this task. Some techniques were developed to help model converge faster, e.g. sortagrad \cite{DeepSpeech2} (using shorter segments at the beginning of training).

To compute CTC-loss we use all possible alignments between audio features and reference text, but only some of the alignments make sense.
Traditional DNN-HMM systems also use iterative training with finding best alignment and then training neural network to approximate this alignment.
Therefore, we propose the following algorithm to use segmentation during training:  
\begin{itemize}
\item compute CTC-alignment (find the sequence of targets with minimal loss that can be mapped to real targets by collapsing repeated characters and removing blanks) 

\item perform greedy decoding (argmax on each step)

\item find ``well-recognized'' words with $length \geq T$ ($T$ is a hyperparameter): segment should start and end with space; word is ``well-recognized'' when argmax decoding is equal to computed alignment

\item if the word is ``well-recognized'', divide the utterance into 5 segments: left segment before space, left space, the word, right space and right segment

\item compute CTC-loss for all this segments separately and do back-propagation as usual
\end{itemize}

The results of training with this criterion are shown in Table \ref{tab:Segmentation}. The proposed criterion doesn't lead to consistent improvement while decoding with large beam width (2000), but shows significant improvement when decoding with smaller beam (100). We plan to further explore utilizing alignment information during training.

\begin{table}
  \caption{Models trained with CTC and proposed CTC modification}
  \label{tab:Segmentation}
  \centering
    \begin{tabular}{ l c c c c c }
    \toprule
    \textbf{Model} & \textbf{Segmentation} & \textbf{Argmax} & \multicolumn{2}{c}{\textbf{LM decoding}} & \textbf{Sclite} \\
    & & & \textbf{beam 100} & \textbf{beam 2000} & \\ 
    \midrule
    DS-SF & - & 68.6 & 58.9 & 54.5 & 49.7 \\
    DS-SF & + & 66.7 & 54.9 & 53.9 & 48.7 \\
    bLSTM & - & 69.0 & 59.6 & 55.7 & 51.1 \\
    bLSTM & + & 70.3 & 58.3 & 56.4 & 51.4 \\
    \bottomrule
    \end{tabular}
\end{table}

\subsection{Using different features}

\begin{table}
  \caption{ 6-layers bLSTM trained using different features and normalization}
  \label{tab:Features}
  \centering
    \begin{tabular}{ l c c c c }
    \toprule
    \textbf{Features} & \textbf{Argmax} & \multicolumn{2}{c}{\textbf{LM decoding}} & \textbf{Sclite} \\
    & & \textbf{beam 100} & \textbf{beam 2000} \\ 
    \midrule
    FBanks & 76.3 & 66.3 & 60.6 & 56.3 \\
    FBanks + CNM & 73.6 & 65.4 & 59.0 & 54.6 \\
    FBanks + CMVN & 74.1 & 64.5 & 59.4 & 55.0 \\
    FBanks + CMN + deltas & 69.0 & 59.6 & \textbf{55.7} & \textbf{51.1} \\
    FBanks + CMVN + deltas & 73.8 & 64.3 & 59.0 & 54.5 \\
    spectrogram & 84.0 & 74.8 & 68.1 & 64.0 \\
    spectrogram + CMN & 74.2 & 63.9 & 59.1 & 54.4 \\
    \bottomrule
    \end{tabular}
\end{table}

We explored different normalization techniques. FBanks with cepstral mean normalization (CMN) perform better than raw FBanks. We found using variance with mean normalization (CMVN) unnecessary for the task. Using deltas and delta-deltas improves model, so we used them in other experiments. 
Models trained with spectrogram features converge slower and to worse minimum, but the difference when using CMN is not very big compared to FBanks.

\subsection{Varying model size and number of layers}

Experiments with varying number of hidden units of 6-layer bLSTM models are presented in Table \ref{tab:Units_Layers}. Models with 512 and 768 hidden units are worse than with 320, but model with 1024 hidden units is significantly better than others. We also observed that model with 6 layers performs better than others.

\begin{table}
  \caption{ Comparison of bLSTM models with different number of hidden units.
 }
  \label{tab:Units_Layers}
  \centering
    \begin{tabular}{ l c c c c c }
    \toprule
    \textbf{Units} & \textbf{Layers} & \textbf{Argmax} & \multicolumn{2}{c}{\textbf{LM decoding}} & \textbf{Sclite} \\
    & & &  \textbf{beam 100} & \textbf{beam 2000} \\ 
    \midrule
    320 & 6 &  69.0 & 59.6 & 55.7 & 51.1 \\
    512 & 6 & 71.4 & 62.2 & 57.1 & 52.5 \\
    768 & 6 & 69.9 & 62.3 & 56.2 & 51.7 \\
    1024 & 6 & 67.3 & 57.0 & \textbf{53.3} & \textbf{48.4} \\
    \midrule
    1024 & 5 & 70.7 & 61.9 & 56.0 & 51.3 \\
    1024 & 7 & 69.3 & 60.6 & 55.9 & 51.4 \\
    \bottomrule
    \end{tabular}
\end{table}

\subsection{Training the best model}

To train our best model we chose the best network from our experiments (6-layer bLSTM with 1024 hidden units), trained it with Adam optimizer and fine-tuned with SGD with momentum using exponential learning rate decay. The best model trained with speed and volume perturbation \cite{Ko15-AAF} achieved 45.8\% WER, which is the best published end-to-end result on Babel Turkish dataset using in-domain data. For comparison, WER of model trained using in-domain data in \cite{Dalmia2018} is 53.1\%, using 4 additional languages (including English Switchboard dataset) – 48.7\%. It is also not far from Kaldi DNN-HMM system \cite{KaldiTurResults} with 43.8\% WER.

\begin{table}
  \caption{ Using data augmentation and finetuning with SGD }
  \label{tab:Finetuning}
  \centering
    \begin{tabular}{ l c c c c }
    \toprule
    \textbf{Augmentation} & \textbf{Argmax} & \multicolumn{2}{c}{\textbf{LM decoding}} & \textbf{Sclite} \\
    & & \textbf{beam 100} & \textbf{beam 2000} \\ 
    \midrule
    speed & 63.8 & 54.0 & 51.0 & 46.2 \\
    speed + volume & 63.5 & 53.6 & 50.7 & \textbf{45.8} \\
    \bottomrule
    \end{tabular}
\end{table}

\section{Conclusions and future work}
In this paper we explored different end-to-end architectures in low-resource ASR task using Babel Turkish dataset. We considered different ways to improve performance and proposed promising CTC-loss modification that uses segmentation during training. Our final system achieved 45.8\% WER using in-domain data only, which is the best published result for Turkish end-to-end systems. Our work also shows than well-tuned end-to-end system can achieve results very close to traditional DNN-HMM systems even for low-resource languages.  
In future work we plan to further investigate different loss modifications (Gram-CTC, ASG) and try to use RNN-Transducers and multi-task learning.

\section{Acknowledgements}
This work was financially supported by the Ministry of Education and Science of the Russian Federation, Contract 14.575.21.0132 (IDRFMEFI57517X0132).

\bibliographystyle{splncs04}


\bibliography{mybib.bib}




\end{document}